\documentclass[12pt]{iopart}

\usepackage{iopams}  
\usepackage{graphicx}
\usepackage{color} 
\usepackage{hyperref}
\usepackage{xcolor}
\usepackage{comment}
\usepackage{slashed} 
\hypersetup{
    colorlinks=true, %set true if you want colored links
    citecolor=black,
}
\usepackage{float}
\usepackage{microtype}\usepackage{placeins}
\begin{document}
%\raggedcolumns

\title[Models and Potentials in Hadron Spectroscopy]{Models and Potentials in Hadron Spectroscopy}

\author{Sreelakshmi M \& Akhilesh Ranjan}

\address{Department of Physics, Manipal Institute of Technology
\\ Manipal 
Academy of Higher Education, Manipal, 576104, Karnataka, India\\}
\ead{ak.ranjan@manipal.edu}
\vspace{10pt}
\begin{indented}
\item[]October 2022
\end{indented}

\begin{abstract}
In the past twenty years, hadron spectroscopy has made immense progress. Experimental facilities have observed different multiquark states during these years. There are different models and phenomenological potentials to study the nature of interquark interaction. In this work, we have reviewed different quark potentials and models used in hadron spectroscopy.
\end{abstract}

\vspace{2pc}
\noindent{\it Keywords}: Quarks, QCD, Hadron spectroscopy, Quark potential
%\maketitle

\section{Introduction}

Quarks, leptons and gauge bosons are considered as the most elementary 
particles observed in laboratories. Due to color confinement, only color 
singlet configurations of quarks are observed in nature. Mesons and baryons
are well known color singlet structures. Further Gell-Mann and Zweig gave the idea of color singlet hadronic state with 
$qq \bar q \bar q$ 
and $qqq q \bar q$ quark combinations known as tetraquarks and pentaquarks 
\cite{gellman, zweig}.

Over the past decades, there have been significant 
advances in experimental facilities. With the help of recent developments 
in high energy experiment and computational techniques, many new hadrons are discovered. Recent developments in lattice gauge theory also support 
experimental results \cite{kronfeld, wingate, s.prelovesek}. After the discovery of 
$J/\Psi$ in 1974, heavy quarkonium studies 
have also become very important in hadron physics \cite{aubert}.

Quantum chromodynamics (QCD) is the theory of strong interaction between quarks and gluons. The study of heavy flavor spectroscopy is important for understanding  
strong interaction. The fully heavy tetraquark and pentaquark states are 
perfect prototypes for improving the knowledge of heavy quark potential. A fully heavy tetraquark or pentaquark state involves the nonperturbative color confinement potential and the perturbative one gluon exchange (OGE) interaction \cite{deng}. %There are many heavy hadrons whose properties are not studied yet. %In knowledge of quark potential, study of various properties of hadrons is important.
 
Since interquark interaction is a non-Abelian and non-linear theory, which is 
very complicated and not understood yet \cite{griffiths}. Therefore, we study 
interquark interactions with some models. None of the models is capable of 
explaining all the hadronic systems. The quantum mechanical potential models can reproduce the experimental results for hadron spectroscopy. Potential 
models are based on the assumption that potential can characterise the 
interaction between the quark and antiquark. When the quark mass is heavy ($m_{Q} >> \Lambda _{QCD}$) and velocity of quark is $v << 1$, the system 
can be treated non relativistically and solving the Schroedinger equation 
will lead to the properties of the system. For solving the Schroedinger 
equation of quark-antiquark potentials, there are several analytical 
techniques. Otherwise we use relativistic potentials.

In this article we shall review the quark potentials and models which are 
extensively used to determine the mass and other properties of hadrons.

\section{Quark potentials and models in hadron spectroscopy}
 There are different methods for hadron spectroscopy to determine the properties
of hadrons, it include the bag model, QCD sum rules, Bethe-Salpeter equation method, various phenomenological potential models, and lattice QCD (LQCD), etc. These methods, with some approximations and assumptions are found to be very 
useful in determining the properties of hadrons.

\subsection{The bag model}
The two important properties of QCD are asymptotic freedom and quark confinement. The origin and nature of quark confinement is still unknown. One of the very 
successful phenomenological model for quark confinement is the bag model 
proposed in 1974 by Chodos {\it et al} \cite{chodos}. 
%The bag model is a 
%relativistic model that can be considered as analog of the nuclear shell 
%model. 

The bag model assumes 
quarks in hadrons are non-interacting and confined in a finite region called 
``bag." The infinite potential of the bag will confine the quarks inside. 
Quarks can move freely inside the bag and they can not reach the exterior due to infinite potential.  

The general structure of Lagrangian density for MIT bag model is,
\begin{equation}
{\mathcal{L}}=({\mathcal{L_{QCD}}}-B)\theta_{v}-\frac{1}{2}\bar q(r)q(r)\Delta_{s},
\end{equation}

where $\theta_{v}$ is the step function which is zero outside and surface of 
the bag and unity inside the space-time region of bag. $B$ is bag constant and 
$\Delta_{s}$ is a function which is unity on the bag surface and zero otherwise.
The other term is a Lagrange multiplier represents the confinement condition.

The bag model provides the mass and other properties of hadrons in 
their ground state \cite{chodos1, degrand, strottman, haxton, aerts}.
Chodos {\it et al} studied baryon structure in static bag model 
\cite{chodos1}. The spectrum for low-lying baryon resonances was obtained by 
assuming the bag as a sphere of constant radius. The Dirac equation was solved by considering quarks as massless. Their study incorporated many successful 
non-relativistic features of the quark model like magnetic moment and gyromagnetic ratio. The effect of quark gluon interaction was neglected here.

DeGrand {\it et al} studied the light hadrons including baryon octet and decuplet, pseudoscalar and vector meson nonets to get the mass and static 
parameters using bag model \cite{degrand}. Unlike the study of Chodos 
{\it et al} \cite{chodos1}, quark-gluon interactions were considered this time. Also, non-strange masses were introduced for quarks. The effect of deviation from the spherical shapes of the bags was also discarded. In this work, the number of free parameters was increased to four. 
Masses of light hadrons for $m_{0}=0$ and $m_{0}=0.108$ $GeV$ (introducing a 
slight non strange mass) were calculated and compared with the experimental 
values. However, the spectrum was insensitive to the non-strange mass 
contribution. The theoretical estimation showed good agreement with the 
experimental results.
Authors had also estimated magnetic moments, charge radii, and weak decay 
constants. 
Their study also showed that exotic hadrons are unstable. At the end of the study, they proposed the possibility of including charm 
quark mass to get the charm hadron spectrum.

After the discovery of the charm quark, Jaffe and Kiskis tried to extend this 
model to get mass spectra of hadrons with charm quarks \cite{jaffe}. They used the cavity approximation to the bag model. Quark-gluon interactions were also taken into account. The mass of hadrons with top $(t)$ and 
bottom $(b)$ quarks were also proposed (these hadrons were new at that time 
because $t$ and $b$ quarks were not confirmed by experiment). The study considered baryons with one heavy quark. The mass predicted for $\Psi^\prime$ was lower than the experimental result. 

Later many improvements were made in the model to get heavy hadron 
properties. Bernotas and  Simonis provided a combined description for light and 
heavy hadrons (mesons and baryons) in the bag model \cite{bernotas}. The center of mass correction was added to heavy hadrons. It helped to obtain mass spectra that were in good agreement with the 
experimental values. Their bag model and the MIT bag model have some 
differences. The zero-point energy term ($Z_{0}$) and self-energy term of MIT bag model were discarded here. The zero point energy term played 
an important role in the MIT bag model to get a good fit. The number of free parameters was further increased. The values of mass of hadrons with heavy quarks showed better results than the MIT bag 
model by Chodos {\it et al} \cite{chodos1}. However, results for light hadrons were not much different. Some of the drawbacks of the model were the difference in the $\pi-K$ mass and $\Sigma_{h}-\Lambda_{h}$ mass splitting. Still, reasonably well results for other hadrons were obtained.

In the MIT bag model combined with chromomagnetic interaction, Zhang, Xu, and Jia calculated the masses and magnetic moments of heavy baryons and tetraquarks with one or two heavy quarks \cite{zhang1}. The results were 
compared with MIT bag model calculations and experimental results. 
They have predicted mass of some states, including the tetraquark state $u d \bar s \bar c$. 

As we have observed, the simple formalism of the bag model allows us to represent mesons, baryons, and exotics. Still, the study of exotics in bag model is significantly less.

\subsection{The Bethe-Salpeter equation (BSE)} 
Salpeter and Bethe developed a mathematical formalism aiming an extension to Feynman's formalism of bound state problems involving several particles 
\cite{salpeter1951}. The original BSE equation used two-body bound states for 
the relativistic interaction kernel and related wave functions. The formula effectively calculates the mass spectra of mesons.

Munczek and Jain discussed the properties of $q \bar q$ pseudoscalar meson 
bound state using the BSE equation coupled with Schwinger-Dyson (SD) equation 
\cite{munczek}. They have used the ladder approximation in the Landau gauge 
theory. Heavy-heavy, heavy-light, and light-light $q \bar q$ states were 
studied. The equations were solved to calculate bound state masses, wave functions, and leptonic 
decay constants. The obtained results and the results from the experiments were in strong agreement. In the case of ladder approximation, the multi-gluon exchange part of the kernel
was neglected, leading to limitations such as color gauge invariance and the absence 
of crossing symmetry. In other work, the same authors have extended their study to obtain the spectrum of vector, scalar, and pseudoscalar meson bound states with heavy and 
light quarks \cite{jain}. They have simplified the model assuming that major contribution comes only from one 
tensor component for a particle with a given spin and parity. The results obtained were similar to the study of Munczek and Jain \cite{munczek}. 

Guleria and Bhatnagar have estimated the mass spectrum and leptonic decay 
constants of heavy-light axial vector mesons ($1^{++}$ and $1^{+-}$) using BSE formalism with ladder approximation \cite{guleria}. The interaction kernel 
consisted of confining term and one gluon exchange (OGE) term. The masses agreed with experimental values and 
results from other methods.

Li, Chang, and Wang used the relativistic Bethe-Salpeter formalism to get the 
mass spectra and wave functions of $b c q $ baryons \cite{li-chang}. 
Using diquark formalism with instantaneous approximation, the three-body 
problem was converted to two two-body problem. The obtained mass spectra of $\Xi_{bc}$ and $\Omega_{bc}$ were consistent with other studies. 
The decay properties for these states were not discussed.

Li {\it et al} studied fully heavy tetraquark states using BSE formalism 
\cite{li}. Using instantaneous approximation, the tetraquark state was taken as a bound state of diquark and an antidiquark. The equation was numerically 
solved to obtain the mass spectra and wave functions. They have used the same model parameters applied for mesons and baryons from Li, Chang, and Wang \cite{li-chang}. The obtained ground state mass spectra for the 
$cc \bar c \bar c$ was $6.4$-$6.5$ $ GeV$. The comparison with other 
approaches was roughly consistent with the ground state of $cc \bar c \bar c$. 
Also, obtained masses were lower than the LHCb obtained mass of $X(6900)$ \cite{lhc6900}. Therefore, they have proposed that the observed $X(6900)$ may not be the 
ground state of $cc \bar c \bar c$ and it may be the first or second radial 
excited state. They suggest a more detailed study of the inner structure of 
this tetraquark state. Masses for the ground states of $bb \bar b \bar b$ were obtained in the 
range of $19.2$-$19.3$ $GeV$. The theoretical values of masses from other studies was higher compared to the obtained results.

Abu-Shady, Gerish and Ahmed studied heavy pentaquarks with $J^{P}$ values 
$\frac{1}{2}^{+}$, $\frac{3}{2}^{+}$, and $\frac{5}{2}^{+}$ in the framework of
spinless Salpeter equation \cite{abu-gerish}. The pentaquark was considered as an antiquark and two diquarks. In this pentaquark at least one of the quark is heavy. By choosing the potential energy of quark interaction as a 
combination of logarithmic potential, linear potential, and spin-dependent 
potential, they have employed the Bethe-Salpeter equation for pentaquarks. 
A logarithmic potential was used for the first time here for pentaquarks. The outcome was comparable with other models. The study did not discuss about decay properties. One gluon exchange approximation using an instantaneous 
potential can only be applied to states containing at least one heavy quark. 
The model parameters were derived using the Cornell and logarithmic potential relationship. 

\subsection{The chromomagnetic interaction model}
Hyperfine structure of hadron spectroscopy involves spin-related interaction between quarks or quarks and antiquarks, which have a color factor. The color-magnetic interaction that results in the mass splitting for ordinary hadrons is caused by the one gluon exchange potential. The Hamiltonian of the color-magnetic interaction, often known as the chromomagnetic interaction (CMI) model is an efficient way to describe hadron masses after the quark mass is included \cite{liu}.

There are many types of CMI Hamiltonian. The general structure of the Hamiltonian for the CMI  model is,

\begin{equation}
    H=\sum_{i}m_{i}-\sum_{i<j}v_{ij}\lambda_{i}\cdot\lambda_{j}\sigma_{i}\cdot\sigma_{j},
 \end{equation}
 
 where $m_{i}$ is the effective mass, $v_{ij}$ is the coupling parameter and 
 $\lambda_{i}$ is the Gell-Mann matrices. CMI models contain coupling coefficients and effective masses as parameters. In a simple CMI model the coupling constant and effective quark mass can be extracted from known hadrons. 
 
In the case of doubly heavy baryons the experimental studies are very few but its theoretical studies are done extensively. Weng, Chen, and Deng studied the masses of doubly heavy and triply heavy baryons using the chromomagnetic 
model with color interaction \cite{weng}. In 2017 LHCb reported the doubly 
charm $\Xi_{cc}^{++}$ state \cite{aaij-d-charm}. The calculated value of 
$\Xi_{cc}$ by this study was close to the LHCb result. For getting the 
model parameters ($m_{qq}$ and $v_{qq}$) of baryons, experimental data of the 
light and singly heavy baryons were used. They had extracted thirteen model 
parameters. The study did not discuss the decay properties. 

Guo {\it et al} studied mass spectra of multi heavy baryons and $S$-wave 
doubly heavy tetraquark $QQ \bar q \bar q$ ($Q=c,b$, $q=u,d,s$) with 
$J^{P}$= $0^{+}$, $1^{+}$, and $2^{+}$ in the improved CMI (ICMI) model, which 
includes chromomagnetic and chromoelectric interactions \cite{guo}. The 
parameters ($m_{ij}$ and $v_{ij}$) were extracted by fitting it with the conventional hadron spectra. Their study included doubly and triply heavy 
baryons. The study also proposed the mass of $ \Xi_{cc}$ baryon. The study gave similar result compared with the study by Weng \cite{weng}. However, they were able to predict the masses of tetraquark states also. The 
predicted mass of $c c \bar u \bar d$ tetraquark state agrees with the LHCb results. The mass of tetraquark states $b b \bar n \bar s$, $bb \bar n \bar n$, $ b c \bar n \bar n$, $bb \bar s \bar s$, $ b c \bar s \bar s$ and $ b c \bar n \bar s$ ($n=u,d$) were also predicted. 

Chen {\it et al} used the simple chromomagnetic interaction model to study 
the triply heavy ($QQ \bar Q\bar q$) tetraquark states \cite{chen}. They have used a diquark-antidiquark $[(QQ)(\bar Q \bar q)]$ approach and a triquark-antiquark $[(QQ \bar Q)( \bar q)]$ approach. Both 
methods gave same results. However, this approach could not predict the masses accurately. The reason comes from the effective coupling constant. Therefore, they suggested an improved model with color-Coulomb term, kinetic term, and confinement term instead of this simple model. They also predicted the decay properties of the states. 

Heavy flavor pentaquark states are also studied using CMI models with 
chromomagnetic and chromoelectric contributions. An {\it et al} studied the 
mass spectra of heavy pentaquarks with four heavy quarks ($QQQQ \bar q$) \cite{an}. They calculated the relative partial decay width of $cccc \bar q $ and 
$bbbb \bar q$ pentaquark states. However, this type of pentaquarks is not 
identified by any experiment till now. Therefore, more investigation on this type of 
pentaquarks states is necessary to identify its exotic nature and other 
properties. An {\it et al} extended their study to the fully heavy pentaquark 
states using the same formalism \cite{an2}. After the systematic calculation of 
the CMI Hamiltonian, mass spectra of $ QQQQ \bar Q$ were calculated.

\subsection{QCD sum rules}
QCD sum rule is an important non-perturbative method developed by Shifman, 
Vainshtein, and Zakharov (SVZ) \cite{shifman}. This is a widely used method 
in hadron phenomenology. This method is extensively used to get the low energy parameters of hadrons. In this method, the time ordered current is expanded into 
a quark and gluon condensate using operator product expansion (OPE) which can parameterize the long-distance attributes of the QCD vacuum. Properties of 
hadrons can be determined from the current-hadron duality. The method is  successfully applied to study the properties of heavy mesons, baryons, and exotics.

Das, Mathur, and Panigrahi investigated the vector meson masses and decay widths using QCD sum rule \cite{das}. The decay properties have been studied 
for the first time with QCD sum rules. They have used the original sum 
rule based only on the two point function of currents given by $\pi_{\mu \nu}$. 
The $\pi_{\mu \nu}$ contains longitudinal and transverse functions. However, 
only the transverse function was considered here. SVZ derived a rule called the Borel-transformed rule for transverse function. This rule was obtained by 
keeping only low dimensional terms in the OPE of the two point function. For masses and decay widths, the experimental value provided the best fit for 
$K^{*}$ and $\phi$. In the case of $\rho$ meson, the calculated width was less than 
the experimental value. This may arise because the four quark condensate contribution was significant and errors can not be neglected. They also found 
that the results were very sensitive to slight variations in the parameters. 

Wang has studied the $\Omega_{b}^{*}$ and $\Omega_{c}^{*}$ ($\frac{3}{2}^{+}$) heavy baryons \cite{wang}. He used the operator product 
expansion assuming vacuum saturation for higher dimension contribution will be suppressed due to a large denominator. Therefore, the contribution from 
the higher dimension condensate was neglected in the calculation. Mass of 
$\Omega_{c}^{*}$ was compatible with the experimental value. For 
$\Omega_{b}^{*}$ the value of masses were compatible with other theoretical 
calculations and lattice QCD values.
 
Exotic states are also successfully studied using QCD sum rules. Zhang has 
performed a study of fully heavy pentaquark ($cccc \bar c$ and $bbbb \bar b$) states \cite{zhang}. 
As in the usual QCD sum rule, two gluon and three gluon condensates were 
considered here. Fully charm pentaquark mass is obtained to be 
$7.41^{+0.27}_{-0.31}$ $GeV$ and for fully bottom pentaquark 
$21.60^{+0.73}_{-0.22}$ $GeV$. However, fully heavy pentaquark states are not 
experimentally detected. They have proposed that this state can be 
searched through $\Omega_{QQQ}\eta_{Q}$ mass spectrum. 

Wang has studied fully heavy hexaquark states using QCD sum rules 
\cite{z.g.wang}. However, there is no experimental evidence on hexaquarks. 
A hexaquark state was considered as three diquarks. They have proposed that the 
hexaquarks can be searched through the $\Omega_{ccc}$ and $\Omega_{bbb}$ invariant mass spectrum. 

In most cases, due to a high degree of accordance with the experimental result, the QCD sum rule prediction is one of the most reliable methods for determining unknown properties of hadrons, particularly heavy hadrons. However, the decay properties are not much explored by the QCD sum rule approach.

\subsection{Diquark model}

The idea of diquark was proposed by Ida and Kobayashi \cite{ida}. The diquark model 
plays an important role in hadron spectroscopy. These are considered as the 
building blocks of exotic hadrons. Diquarks are tightly bound 
colored objects with two possible $SU(3)$ representations. The direct product 
of diquark results in a color antitriplet and a color sextet, 
$3\otimes3=\bar 3\oplus6$.

The product of $SU(3)$ matrices contain,
\begin{equation}
    t_{ij}^{a} \cdot t_{kl}^{a}=\frac{-1}{3}(\delta_{ij}\delta_{kl}-\delta_{il}\delta_{kj})+\frac{1}{6}(\delta_{ij}\delta_{kl}+\delta_{il}\delta_{kj}),
\end{equation}

where the first term represents the antisymmetric product with a negative coefficient and the symmetric term has a positive coefficient reflecting the repulsion. Therefore a diquark is assumed to have $SU(3)$ antitriplet with the antidiquark, a color triplet. Jaffe studied the diquark correlation in QCD \cite{jaffe1}. It is believed that the diquark correlation can give an answer to some questions in 
exotic hadron spectroscopy including the rarity of exotics in QCD. Diquarks are spin $0$ or spin $1$ system.

Anwar, Feretti, and Santopinto calculated the spectrum of hidden charm 
tetraquark states $q c \bar q \bar c$ and $ s c  \bar s \bar c$ \cite{anwar}. 
The spectrum of tetraquarks was obtained in two steps. Using a relativized 
quark-quark potential, the diquark masses were obtained. Then tetraquark spectrum was calculated using the relativized diquark-antidiquark potential. The 
relativistic potential contains one gluon exchange plus confining potential. 
With $qq \bar c \bar c$ quark assignment, they obtained tetraquark states $X(3872)$, $Z_{c}(3900)$, $Z_{c}(4020)$, $Y(4008)$, $Z_{c}(4240)$,
$Y (4260)$, $Y (4360)$, and $Y (4660)$. With $s \bar c s \bar c$, they obtained 
tetraquark states $X(4140)$, $Y(4500)$ and $X (4700)$. 

Exotics can be assumed as diquark states and different models can be applied 
to the system. Some of these are already discussed in the previous and upcoming 
models.

\subsection{String models}
String model is an interesting model in hadron physics \cite{nambu}. For large 
interquark separations, QCD perturbation method fails 
and there is confinement of quarks. The color confinement is an intrinsically 
nonperturbative phenomena. String model is one of the method which explains this behaviour. Still it is not a complete picture.

Mesons are considered as a string of quarks and antiquarks. The site where a quark-antiquark is linked will be a color singlet. When 
a quark and an antiquark are getting far apart more strings have to be excited 
to connect the two sites. When the energy is enough to create new hadrons, system breaks and new pair forms \cite{riazuddin}. The string has linearly 
varying energy and constant energy density. This is called the meson string 
model. 

\begin{equation}
    V(r) \sim a r,
\end{equation}

where $a$ is a constant. Artru considered a string with quark at one end and an antiquark at other end 
for mesons and baryons with three strings joining at a point with quarks at 
free ends. The study allowed construction of exotic states \cite{artru}. The 
string picture clearly tells the absence of free quarks. Quark spin and quark 
statistics are not considered in this work.

String models of baryons can be 
$q-q-q$ configuration, three string modes or $Y$ configuration and triangle model or $\Delta$ configuration proposed by Sharov \cite{sharov, sharov1, sharov2}. The study says the linear string model of baryons is unstable for any value of 
mass \cite{sharov}. Sharov has compared the linear 
string model of baryons with the $Y$ configuration string model of baryons 
\cite{sharov1}. One drawback of this model was the value predicted for the 
slope ($\alpha^{'}$) of the Regge trajectories ($J\sim\alpha^{'}E^{2}$) 
differs from the value of $\alpha^{'}$ of mesons by a factor of $2/3$ at large 
$E$ values.

There are different variants of the string model. Olsson string model approached 
mesons using various equations like the Bethe-Salpeter equation and generalized 
Klein-Gordon equation. Goebel, LaCourse, and Olsson explained the Regge 
trajectories with the help of the string model and showed that on varying the quark 
mass and considering the Coulomb interaction could violate the linear nature of 
Regge trajectories \cite{goebel}. The study concludes that when the quarks are massless, both vector and 
linear confinement results in parallel Regge trajectories. When a potential deviates from linear confinement at small 
radii, it can result in linear Regge trajectories at higher angular momenta. 
Another type of string model was given by Soloviev \cite{soloviev}. Here the author had 
considered a relativistic spinning model of quarks and antiquarks (massive), which led to Regge trajectories of mesons. Current quark masses and string tensions were the main 
parameters here. The model prediction was compared with the experimentally identified meson 
masses to get the current quark masses.

\subsubsection{Regge trajectories}

Regge trajectories of hadrons were introduced by T. Regge in 1959 \cite{regge1, 
regge2}. The mass and angular momentum of hadrons are related by the equation \cite{mike},
\begin{equation}
m^{2}_{J,n}=aJ+bn+c,
\end{equation}

where $n$ is the principal quantum number and $a$, $b$, and $c$ are the slope 
and intercept parameters. The hadrons lie on a line in the $(J,M_{2})$ plane called Regge trajectories. Usually string picture of hadrons are the most 
popular model used to explain the Regge trajectories. Regge trajectory can be 
used to obtain the mass spectra of hadrons. As explained earlier in string 
models by Goebel, LaCourse, Olsson, and Soloviev formation of Regge 
trajectories were discussed \cite{goebel, soloviev}. 

A study by Semay and Brac used the relativistic flux tube model and obtained a linear Regge trajectory for mesons in the ultra-relativistic limit \cite{semay}. 
The Coulomb-like potential, instanton 
effect, and a constant potential lead to the flux tube picture. The constant potential was taken as a 
negative value to reproduce the data. The spin effect is added to the potential with the help of the instanton interaction. Here, they assumed the instanton effect acts only for $L=0$ meson state. 
The study showed that the nature of constant potential is important in 
getting good results. Satisfactory results for many meson spectra were 
obtained. The constant potential coming from the extremities of the flux 
tube has given the most satisfying results with theories and experiments.

Sergeenko combined heavy and light quarkonia to develop an analytic 
expression \cite{sergeenko}. Regge trajectories of heavy and light quarkonia in all regions were developed. They used Cornell potential 
with a constant term. For the square of mass, an analytical equation was developed. The formula incorporates spin-orbit and spin-spin interactions. The interpolating mass formula 
showed good accuracy in obtaining spectra of bound states of quarkonia. 

Ebert, Faustov, and Galkin used the relativistic quark model to study 
the mass and Regge trajectories of light mesons using a quasipotential \cite{ebert}. The quasipotential was derived by assuming the resultant 
interaction was the combination of one gluon exchange with long-range scalar and vector linear confining potential. They were able to calculate masses and the linear Regge trajectories of light mesons. 
 
Nandan and Ranjan investigated the Regge trajectories of pentaquarks with 
different possible configurations using the flux tube model \cite{ranjan}. 
Regge trajectories for pentaquarks showed deviation from linear behavior. The 
result was compared with the available experimental results. At low rotational speed, the mass of pentaquark showed a linear increase, but at 
the high rotational speed of string, the Regge trajectories became highly 
nonlinear. They also observed that two different pentaquarks could show the same 
mass and angular momentum.
 
It can be clearly seen that, more studies are needed in the case of Regge 
trajectories of exotic candidates.

\subsection{Skyrme model and chiral quark-soliton  ($\chi$QSM) model}

This model was given by Skyrme in $1961$ \cite{skyrme}. The baryons emerged as collective excitations of mesons fields. This model is widely used in 
hadron physics. The model approaches to $\chi$QSM model in the large $N_{c}$ limit. Skyrme introduced a Lagrangian density by adding 
non linear-$\sigma$ model Lagrangian into it \cite{callan},
\[{\mathcal{L}}_{Skyrme}=\frac{f_{\pi^{2}}}{16}\Tr(\partial_{\mu}U\partial^{\mu}U^{\dagger})+\frac{1}{32e^{2}}\Tr\{[U^{\dagger}\partial_{\mu}U,U^{\dagger}\partial_{\nu}U][U^{\dagger}\partial^{\mu}U,U^{\dagger}\partial^{\nu}U]\},\]

where $e$ is a dimensionless parameter, $U$ is the field. From the Lagrangian, 
energy can be calculated and mass can be obtained. In some cases, the Skyrme Lagrangian also includes symmetry-breaking terms. Praszalowicz studied 
pentaquaks in Skyrme model \cite{praszalowicz}. He has shown that 
the chiral model also predicts existence of low-lying exotic baryons. They estimated the mass of $\theta^{+}$ pentaquark state. 
However, $\theta^{+}$ state is not yet confirmed by experiments. 

Callan and Klebanov developed a mass relation for strange baryons using this model \cite{callan}. With $SU(3)$ symmetry, the strange baryons were treated as the bound state of kaons. The rotation of the skyrmion was 
ignored to get baryon masses to $O({N_{c}^{0}})$. The terms in the kaon fields up to second-order were considered and higher-order terms were neglected because 
they represent the self-interaction of kaons. 

The chiral quark soliton model and Skyrme model has similar group structure. 
In the large $N_{c}$ limit, baryons emerge as solitons due to chiral action. 
This model based on baryons is called the chiral quark soliton model. Yang 
and Kim derived mass splitting of $SU(3)$ baryons using the chiral quark 
soliton model by considering the isospin symmetry breaking \cite{yang}. They 
have also developed a model independent approach to derive mass relations. 
Different mass relations were derived for the baryon decuplet, antidecuplet, and octet. They obtained the Gell-Mann-Okubo mass relation, which 
agreed well with the experiment. The known experimental results for baryon decuplets determine the unknown model parameters. Masses for $\Sigma^{*}$ 
and $\Xi^{*}$ have shown good agreement with the experiment. Compared with the previous studies, the second moment of inertia helped to explain the heavier baryon antidecuplet masses, which were difficult to fix 
in previous cases. In the present work authors obtained the masses of baryon antidecuplet 
and decuplet but did not consider the decay widths.

\subsection{Hadroquarkonium model}

This study is inspired by the analogy of hydrogen atom \cite{ali2017}. Surrounding a light matter, a $Q \bar Q$ pair ($Q = c, b$) forms a hardcore. 
The light matter will be $q \bar q$ for tetraquarks and $qqq$ for pentaquarks. In another way, a pentaquark can be considered as a baryon-bound state and an 
excited quarkonium state. The effective Hamiltonian for QCD multipole expansion can be stated as,

\begin{equation}
    H_{eff}=\frac{-1}{2} \alpha^{(\Psi_{1}\Psi_{2})}E_{i}^{a}E_{j}^{a},
\end{equation}

where $E_{i}^{a}$ represents a chromoelectric field and  
$\alpha^{(\Psi_{1}\Psi_{2})}$ is the chromopolarisability. 
Eides, Petrov, and Polyakov considered hidden charm pentaquarks as 
hadroquarkonium states in a QCD motivated method \cite{eides}. 
They have studied the pentaquark $P_{c}(4450)$ \cite{aaij} as a bound state of 
$\Psi^{'}$-nucleon. They calculated the decay width for the state and it agreed with 
the experimental data. This pentaquark was expected to be one of the hidden charm 
baryon octet members. Masses for the octet pentaquarks were also calculated. 
However, the hadrocharmonium approach could not explain the $P_{c}(4380)$ 
pentaquark.

\subsection{One pion exchange potential (OPEP)}
A nucleon can be considered as a bound state formed due to the exchange of 
meson, mainly by pion proposed by Yukawa. The long range pion exchange 
potential is,

\begin{equation}
    V(r)=\frac{f^{2}}{3}\left[(\sigma_{1}\cdot \sigma_{2})(\tau_{1} \cdot \tau_{2})+S_{12}(\tau_{1} \cdot \tau_{2})\left(1+\frac{3}{\mu r}+\frac{3}{(\mu r)^{2}}\right)\right]\frac{e^{-\mu r}}{r}.
\end{equation}

The mechanism of one pion exchange is used to explain the tetraquark with 
a hidden charm which was able to predict some new tetraquark states 
\cite{toernqvist1, toernqvist2}. Toernqvist tried to explain some 
mesonic states which were not fitting with the conventional mesonic 
$q \bar q$ model. He suggested the term ``deuson" for deuteron like meson-meson 
state \cite{toernqvist2}. He expected that the pion exchange would play a 
prominent role here. Toernqvist explained the $X(3872)$ as a $D \bar D^{*}$ 
deuson and they have predicted the masses of these deusons only from pion 
exchange contribution \cite{toernqvist3}. The mass of $D \bar D^{*}$ with $J^{PC}$ 
value $1^{++}$ and $0^{-+}$ were obtained as $3870$ $MeV$, which matched with the 
$X(3872)$ state. This was experimentally confirmed by Belle later \cite{choi}. 

Eides, Petrov and Polyakov used these concepts for pentaquarks \cite{eides}. The $P_{c}(4450)$ pentaquark was considered as a bound state of 
$\Sigma_{c} \bar D^{*}$. The $P_{c}(4380)$ pentaquark could not get a satisfactory result by this explanation. They have also found difficulty in 
explaining the decay of $P_{c}(4450)$ pentaquark.

\subsection{Quark pair creation model}
 The decay of hadrons discusses the creation of quark-antiquark pairs which 
 depends upon QCD. Since there are difficulties in understanding 
 nonperturbative QCD, quark pair creation model is adopted.

 \subsubsection{$P_{0}^{3}$ model or TPZ model}
  Micu proposed the model that discussed the decays of meson resonance \cite{micu}. According to the model, each hadron undergoes a decay that produces a quark pair with the quantum number $0^{++}$ from vacuum, which is then mixed with the quark pairs from the parent hadrons to create the daughter hadrons. The model was used to obtain the strong decays of $2^{+}$, $1^{+}$, and $0^{+}$ meson. Later in 1970s, this model was further modified by Orsay group 
 \cite{leya1,leya2}. This model is now extensively applied on hadron strong decays. 
 Following the production of the quark pair, the initial quarks are rearranged 
 to produce an initial mock state. The decay amplitude was obtained by combining the overlap integrals in spin, color, flavor, and orbital spaces for the 
 initial state and the final state. 
 
 In a separate work, Feng {\it et al} used this model to get the decay width 
 of $S_{1}^{3}$ mesons \cite{feng}. Mass calculated from the Regge trajectories 
 was compared with the experimental mass to identify the possible candidates. Then using $P_{0}^{3}$ model decay widths were calculated. One of the two model parameters, represents effective radius of the particle. They compared the theoretical values with the experimental 
 values. Mass and width of $\rho(1900)$ and $\omega(1960)$ states matched with 
 experimental values. For $\phi(2170)$ the calculated value and the experimental value for partial and total width showed a mismatch. $K^{*}(1410)$ state also contradicted with the experimental value in both mass and width.

\section{Phenomenological potential models}
Researchers are investigating several types of potential models to find out the nature of interquark interaction. Nowadays many studies are done by combining two or more potentials. Some of them are Coulomb and linear potentials, exponential-type potentials, power potentials, harmonic and anharmonic potentials, etc.

\subsection{The Coulomb potential}
  In hydrogen atom, we consider the potential energy of the form \cite{griffith},
\[V(r)=\frac{-e^{2}}{4\pi\epsilon_{0}}\frac{1}{r},\] 

known as the Coulomb potential. Coulomb potential is a very important potential for quark interaction. All the known 
hadrons are color singlets and the gluon exchange between the color singlet 
states provides binding energy between quarks. The two body gluon exchange 
potential has the form \cite{riazuddin},
  \[V_{ij}=\frac{k_{s}\alpha_{s}}{r},\]
  
where $\alpha_{s}$ is the running coupling constant. For short distances the 
QCD perturbation theory can also be applied and one gluon exchange potential can also be considered.

\subsection{The Cornell potential}
 Cornell potential is one of the old and extensively used QCD motivated 
potential developed by the Cornell 
University group. The potential is mainly used for getting the mass spectrum of heavy quarkonia \cite{eichten, kuchin, ghalenovi}. The potential is a combination of Coulomb and the confining term and has been extensively investigated. These terms may have different functional forms. One of the most extensively used forms of potential is,
\begin{equation}
    V(r)= \frac{-a}{r}+br,
\end{equation}

where $a$ and $b$ are positive parameters and $r$ is the interquark distance. 
The potential is quite helpful in addressing the nature of the QCD vacuum, which is paramagnetic and dielectric \cite{cheng}. 

Kuchin and Maksimenko obtained an analytical solution for Cornell potential by applying the Nikiforov-Uvarov (NU) method and studied the mass spectrum of charmonium, 
bottomonium, and $B_{c}$ mesons \cite{kuchin}. They modified  the variable 
as $x=\frac{1}{r}$ and proposed some approximation scheme in the term $\frac{a}{x}$. They have assumed a characteristic radius $r_{0}$ for meson, 
and $\frac{a}{x}$ was expanded in a power series around $r_{0}$ till the second 
order. This helped to deform the centrifugal potential and this modification 
was solved by the NU method. The results showed a good agreement with the experimental values and other theoretical values. 

Lahkar, Choudhury, and Hazarika used Cornell potential to analyze the mass and decay properties of 
heavy flavor mesons containing one heavy quark or antiquark with the help of the variational method using Coulomb, Gaussian, and Airy trial wave functions \cite{lahkar}. Properties like mass, decay constant, branching ratios of leptonic decays, and oscillation frequency of 
heavy mesons were determined. They compared the results with experimental, lattice, and QCD sum rule studies. 

Vega and Flores used Cornell potential to describe properties of $c \bar c$,  
$b \bar b$, and $b \bar c$ states \cite{vega}. The Shroedinger equation was 
solved using an approximate method involving variational method and 
supersymmetric quantum mechanics (SUSY QM). The variational method is an effective tool to get the approximate ground state energies. Supersymmetric QM will help to get the solution of higher energy states. The energy and wave functions of $s$ state of $c \bar c$, $ b\bar b$, and $b \bar c$ were obtained. They calculated the energies for the first three states. Approximate wave functions for the ground, first, and second excited states were also provided. The calculated value showed good agreement with the computational value for the ground and first excited states. The decay properties and experimental values agreed equally well.

The charmonium mass spectra and decay properties were presented using Cornell potential by Chaturvedi and Rai \cite{chathurvedi}. Relative corrections were added to the 
Cornell and spin-dependent potential terms. They successfully determined the mass spectra and decay properties of 
charmonium. The Regge trajectories for charmonium states were also determined, 
which helped to associate some higher excited states with charmonium. 
Calculated charmonium masses fit the Regge planes perfectly. All the results were compared with experimental results. For $1S$ and $2S$ states, the splitting of the energy level was found to be 
higher than the experimental value. For $3S$ and $4S$ states, the results were 
in better agreement with the results from experiments and other theoretical studies.
They have associated the $X(3915)$ and $X(3872)$ as $2^{0}P_{3}$ and  
$2^{1}P_{3}$ respectively. 

Patel, Shah, and Vinodkumar studied masses of hidden charm tetraquark state $cq \bar c \bar q$ $(q \in u, d)$ using Cornell potential \cite{s.patel}. 
Spin-dependent interactions were added to the potential including spin-orbit, spin-spin, and tensor terms. The model parameters include constituent quark 
masses and string tension, which were chosen to fit with the ground state 
masses of experimentally observed $X(3823)$, $Z_{c}(3900)$, and $Z_{c}(3885)$ states. The four body system was considered as two two-body system, with one as a combination of diquark-antidiquark and the other as a cluster of quark-antiquark. Here they assume one quark moves in a static potential of other quarks like the 
hydrogen atom problem. The tetraquark states $X(3823)$, $X(3915)$, $X(4160)$, $Z_{c}(3900)$, $Z_{c}(4025)$, and $\Psi (4040)$ was interpreted as cluster of quark and antiquark. The tetraquark states $X(3940)$, $Y(4140)$, and $Z_{c}(3885)$ were fitted with the diquark-antidiquark 
formalism. The $X(3940)$ as diquark-antidiquark with $2^{++}$ gives good agreement with 
experiment and study by Mainani {\it et al} \cite{mainai}, whereas $Z_{1}(4050)$ agrees more with the formalism of cluster of 
quark-antiquark. 

\subsection{The Martin potential}
Martin potential has the form \cite{martin},
\begin{equation}
    V(r)= a+br^{\alpha},
\end{equation}

where $\alpha \sim 0.1$. Their study was motivated by the success of potential 
models found in the case of heavy quarkonia. The potential generates all the known levels of $J/ \Psi$ and $\Upsilon$ systems studied by Martin 
previously \cite{martin1}. One motivation for the study was realising the need for relativistic correction to the heavy quarkonium $c \bar c$ 
and $b \bar b$. Inspired by the accuracy of the fit, they considered an assumption that a Fermi-type term is controlling the hyperfine splitting. 
For $c \bar c$ system $1S$, $2S$, and $3S$ states lead to the value of $\alpha$ to be $0.1$ with good accuracy. They calculated the masses and relative leptonic decay 
width of $c \bar c$, $b \bar b$ and $s \bar s$. They could only calculate the absolute decay width of $\phi$ from $J/\Psi$. The leptonic decay width of $\phi$ was obtained as $1.6\pm0.23$ $KeV$ and the 
experimental result for the same was $1.43\pm0.12$ $KeV$. However, this is 
accepted because leptonic decay width is sensitive to the wave function and 
relativistic effects compared to the energy levels. Prediction of masses for these states also agreed with the experimental result. 

\subsection{The Coulomb plus power potential}
Patel and Vinodkumar studied the $Q \bar Q$ 
system using the Coulomb plus power potential $(CPP)_{\nu}$ using different values of $\nu$ \cite{patel},
\begin{equation}
   V(r)=-\frac{\alpha_{c}}{r}+ar^{\nu}.
 \end{equation}
 
This potential is a part of the general form \cite{ikhdair, motyka},
\begin{equation}
    V(r)=-cr^{\alpha}+dr^{\beta}+V_{0},
\end{equation}

where $\alpha=-1$, $\beta=\nu$ and $V_{0}=0$. The study mainly used the power range $0.1 < \nu < 2.0$. 

Different potential form will result from different $\nu$ values. The inappropriate choice of the radial wave function for 
heavy quarkonia can affect the decay width and spin splitting of $J$ values 
because both are dependent on the radial wave function of $Q \bar Q$. Therefore, the value of $A$ was 
limited as slightly varying according to the principal quantum number $n$. The 
Shroedinger equation was solved by the method given by Lucha and Schoeberl \cite{lucha}. This potential gave mass spectrum of 
$c \bar c$, $b \bar c$, and $ b \bar b$ mesons up to few excited states. The 
excited states with $\nu=0.9$ to $1.3$ agreed with the results from experiments and other theoretical predictions. They also determined the decay constants of $1S$ to $6S$ states with and 
without QCD corrections. Without QCD corrections, the value of the decay constants for $c \bar c$ systems coincides very well with the experimental value, but not for $b \bar b$ systems. For the $b \bar c$ system, a comparison was not possible at that time due to the unavailability of experimental details. The 
di-gamma and leptonic decay widths were calculated. Di-gamma decay widths agree with experimental values for the range $1.1$ to $1.3$. However, in the same range leptonic decay widths were overestimated or underestimated with radiative corrections. They expect it may be because the decay occurs at some finite separation, not at zero separation.

\subsection{Power-law potential}
Ciftci and Koru used power-law potential to study the leptonic decay widths and 
decay constants of mesons \cite{ciftci}. The potential considered was of the form,

\begin{equation}
    V(r)=\frac{1}{2}(1+\beta)(Ar^{\nu}+V_{0}),
\end{equation}
where $A$ and $\nu$ were greater than zero. They solved the Dirac equation with variation technique. Potential parameters were chosen 
as $A=0.68$ $GeV$, $V_{0}=-0.3961$ $GeV$, and $\nu=0.2$. When spin-spin and hyperfine 
interactions were added to the potential they could obtain good results for 
meson mass spectra.

\subsection{Potential by Jena, Behera, and Panda}
Jena, Behera, and Panda used the assumption that a quark and antiquark in a meson are confined in a potential of the form \cite{jena, jena1998, jena2002},

\begin{equation}
     V(r)=\frac{1}{2}(1+\gamma_{0})(a^{2}r+V_{0}),
 \end{equation}
 
 where $a>0$. The model parameters were $a$, $V_{0}$, and nonstrange quark mass. 
 This model considers the spin-dependent forces as perturbation resulting from 
 the one gluon exchange. The effect of the center of mass motion was also 
 considered. They considered the quark-Lagrangian density for this model in zeroth order as,

 \begin{equation}
    \mathcal{L}_{q}^{0}(x)=\bar \Psi_{q}(x)\left[\frac{i}{2}\gamma^{\mu}\partial_{\mu}-m_{q}-V_{q}(r)\right]\Psi_{q}(x).
  \end{equation}
  
 The mass and decay constant of the pion and the masses of the $\rho$ and $\omega$-mesons were calculated perturbatively. The value of the decay constant agreed with the experimental results.
 
 Later Jena, Behera, and Tripathy extended the study to the radiative transition
 of light and heavy flavor mesons \cite{jena2002}. Model parameters were preserved as the same. They have included the momentum dependence due to the 
 recoiling of the daughter meson. They found improvement in $M1$ transition for 
 light mesons. The heavy meson decay width was also comparable with other results.
 
\subsection{The Song and Lin potential}
Song and Lin proposed a potential for the heavy quarkonium of 
the form \cite{xiaotong},

\begin{equation}
    V(r)=ar^{\frac{1}{2}}-br^{-\frac{1}{2}}.
\end{equation}

The potential is a mixture of inverse square root and square root terms. The 
factors $a$ and $b$ were adjustable parameters. Relativistic effects were also included using spin-orbit and spin-spin terms. The Shroedinger equation was solved using the numerical method. Spin-dependent part of the potential was taken as perturbation. The numerical values for the calculated energy levels of $c \bar c$ and $b \bar b$ did not show much difference when compared with other potential model approaches because $r$ does not vary much. However, it shows clear differences in the case of $t \bar t$ for very small distances ($r<0.1$ $fm$). The energy levels of $c \bar c$, $b \bar b$, and $t \bar t$ and the decay rates of $c \bar c$, $b \bar b$, 
$t \bar t$, and $s \bar s$ were calculated. Most of the results showed 
improvement compared with experiments and other potential model studies.

\subsection{The Turin potential}
Lichtenberg  \textit{et al} proposed a new phenomenological potential which lies between Cornell and Song-Lin potential \cite{litchen},
\begin{equation}
    V(r)=-ar^{-\frac{3}{4}}+br^{\frac{3}{4}}+c.
\end{equation}
The study by Lichtenberg restricted only to the bottomonium case because it is the least relativistic case among quarkonium. Because a study by Jacobs 
{\it et al} suggests relativistic corrections in bottomonium are minimal \cite{jacobs}. They have included higher energy levels assuming that the higher energy levels 
can give more knowledge about the long-range part of the potential. However, the decay rates were not considered for the study. They have done a comparative study using different potentials, including Indiana potential (which will be discussed later), Martin potential, 
Cornell potential, Song-Lin potential and Turin potential (the name given in \cite{ikhdair}). Turin potential was applied for the first time for the quarkonium system. They have assumed that the bottomonium interaction in these static potentials depends only on the distance between the particles ($r$). The energy levels for bottomonium were calculated using all these potentials. 
It fitted equally well for Cornell, Song-Lin, and Turin potential when $b$ quark
mass varied appropriately and with the vanishing constant term $c$. They also compared the energy level differences and spin averaged energy levels 
for these potentials. They have obtained the $\chi^{2}$ value by fitting the 
energy differences. 
 
\subsection{The Harmonic oscillator potential}
It is a very important potential in quantum mechanics because it is one of the few quantum mechanical systems for which there is a precise analytical solution. A harmonic oscillator potential takes the form \cite{griffith},

\begin{equation}
    V(r)=kr^{2},
\end{equation}
 
 where $k$ is force constant. Harmonic oscillator potential is an important term in many of the 
 phenomenological potentials like Killingbeck potential.

Mansour and Gamal chose harmonic oscillator potential with linear and Yukawa 
potentials to study the mass spectra of quarkonium systems $c\bar c$, $b \bar b$ and $\bar b c$ using Nikiforov-Uvarov method \cite{mansour1}. They have also 
added the relativistic corrections with spin-spin, spin-orbit, and tensor 
interactions. The exponential term was expanded by Taylor series up to second 
order. Their results were in good agreement with experimental data.
   
 \subsection{The Killingbeck potential}
  The Killingbeck potential has the form,
  \begin{equation}
       V(r)=ar^{2}+br-\frac{c}{r},
   \end{equation}
 and the potential is known as extended Cornell potential when an inverse 
quadratic term is added to it. 

Abu-Shady {\it et al} studied the heavy meson system using the extended Cornell potential \cite{abu}. Here the N-dimensional Shroedinger equation was 
analytically solved by the Nikiforov-Uvarov method for $N=3$. The obtained results were 
applied to 
$c \bar c$, $b \bar b$, $b \bar c$, and $c \bar s$ system to get the mass 
spectra. The parameters in the model were chosen to fit the 
experimental data. The energy eigenfunctions and eigenvalues were also obtained in higher-dimensional space. For $N=3$, heavy meson masses were obtained. The reported results showed well agreement with experimental result. In the 
case of $b \bar c$ mesons, enough experimental data was not available. In the 
case of $c \bar s$ mesons $1S$, $2S$ and $1D$ states were found close to the experimental value.

Omugbe {\it et al} have also studied the non symmetric extended
Cornell potential to get the mass spectra of $b \bar b$, $c \bar c$, $b \bar c$, and $c \bar s$ system \cite{omugbe}. The addition of the harmonic oscillator potential 
and inverse quadratic potential modifies the behavior when $r\rightarrow0$. 
The problem was solved using the WKB framework. The findings of this study were consistent with those of other analytical techniques and published experimental data.

Salehi investigated ground and excited states of some baryons including $N$, 
$\Delta$, $\Sigma$, $\Xi$, $\Omega$ using the Killingbeck plus isotonic 
oscillator potential of the form \cite{salehi}, 

\begin{equation}
    V(r)=ar^{2}+br+\frac{c}{r}++\frac{d}{r^{2}}+\frac{hr}{r^{2}+1}+\frac{kr^{2}}{(r^{2}+1)^{2}}.
\end{equation}

The Schroedinger equation was solved numerically to get energy eigenvalues. To obtain the baryon energies and determine the baryon masses, the values were fitted using the parameters of the generalised Gursey-Radicati mass formula. The potential model agreed well with the spectrum of octets and decuplets.

\subsection{The Polynomial potential}
Mansour and Gamal used the polynomial potential to obtain the mass spectra $c \bar c$, $b \bar b$, and $B_{c}$ mesons  \cite{mansour2}. 
The mathematical form of the potential is,
\begin{equation}
    V(r)=\sum_{m=0}^{m}A_{m-2}r^{m-2}, m=0,1,2...
\end{equation}
A special case of this potential used in the study was,
\begin{equation}
    V(r)=\frac{b}{r}+ar+dr^{2}+pr^{4}.
\end{equation}

The Schroedinger equation was solved using Nikiforov-Uvarov 
method to get the energy eigenstates. Here the first term corresponds to Coulomb potential. The term linear in $r$ 
represents that $V(r)$ is continuously growing as $r \rightarrow \infty$ and 
leads to quark confinement. The third and fourth terms are harmonic and anharmonic terms responsible for quark confinement. The mass spectra for these three states were studied. The study also showed that the harmonic term 
gives more accuracy to results compared to other potentials.

\subsection{The Kratzer potential}
Another potential form mainly used to study heavy quarkonia is the Kratzer 
potential. It is an extensively used potential to study molecular structure and 
interactions. The form of the potential is \cite{hassan},
\begin{equation}
   V(r)= \frac{a}{r}+\frac{b}{r^{2}}.
\end{equation}
In molecular physics this potential is often written as \cite{aygun},

\begin{equation}
     V(r)= -2D_{e}\left(\frac{a}{r}-\frac{a^{2}}{2r^{2}}\right),
\end{equation}
where $a$ is the internuclear separation and $D_{e}$ is the dissociation energy. The study by Bayrak {\it et al} presented a method for 
calculating the solution for non-zero angular momentum state by Kratzer 
potential using the asymptotic iteration method \cite{byrak}. Kratzer potential can be combined with other potentials to solve the mass spectra of heavy quarkonia.

Inyang {\it et al} used Kratzer potential mixed with the 
screened Coulomb potential to solve the mass spectra of charmonium and 
bottomonium states \cite{inyang}. The series expansion method was used to get the solution. 
The model parameters for charmonium were calculated by solving the two algebraic 
equations using the experimental result for the $2S$ and $2P$ states. Similarly, 
experimental value of $1S$ and $2S$ states was used for bottomonium. They have 
applied their results to calculate the masses for $1S$, $2S$, $1P$, $2P$, $3S$, $4S$, $1D$, 
and $2D$ states. The findings were quite well in agreement with those of the experiments and other theoretical investigations.

\subsection{Cornell, Gaussian, and inverse square potential}
Moazami, Hassanabadi, and Zarrinkamar gave a nonrelativistic potential to get 
the mass spectrum of heavy-light mesons \cite{moazami}. The proposed potential 
takes the form,

\begin{equation}
    V(r)=\frac{a}{r}+\frac{b}{r^{2}}+k_{0}e^{-\frac{\alpha^{2} r^{2}}{2}}+cr,
\end{equation}

where $a$, $b$, $c$, $k_{0}$, and $\alpha$ are constants. The model studied the 
$S$ and $P$ states of $B$, $B_{s}$, $D$, and $D_{s}$ mesons. They have solved the 
Schroedinger equation by considering the inverse square term and Gaussian term as perturbation. The unperturbed part was solved using the 
Nikiforov-Uvarov method. They have obtained mass spectrum, decay constants, 
leptonic decay width, and semileptonic decay width for these mesons. The value of mass obtained was compared with other models and most of the values were in 
good agreement.

\subsection{The Wisconsin potential}
This is a QCD motivated potential. The potential shows perturbative QCD characteristics at close range and linear confinement characteristics at far range \cite{wisconsin}. 
The potential has the form,
\begin{equation}
    V_{W}(r)=V_{I}(r)+V_{s}(r)+V_{L}(r),
\end{equation}

where $V_{s}(r)$ is a short range potential which is regularized two-loop perturbative potential. $V_{I}(r)$ is an intermediate potential which has the form  $V_{I}(r)=r(c_{1}+c_{2}r)e^{-\frac{r}{r_{0}}}$ vanishing for small and 
large quark separations. $V_{L}(r)=ar$ is a long range potential which represents quark 
confinement. 

Jacobs, Olsson, and Suchyta proposed a method to get the solution of the 
Schroedinger and spinless Salpeter equations with QCD inspired Cornell and Wisconsin potential \cite{jacobs}. The potential 
parameters and quark masses were varied to get good agreement with the 
experimental data for both Shroedinger and spinless Salpeter equations. 
They have obtained the charmonium and bottomonium energy levels. The ratios of charmonium and bottomonium energy levels, which were already satisfactory and found to be slightly improved by using relativistic kinetic energy and wave function corrections. The Wisconsin potential showed good results compared to the Cornell potential. 
 
 \subsection{The Yukawa potential or the screened Coulomb potential}
There are different forms of exponential type potential. Exponential potentials are important in nuclear physics, including the Woods-Saxon (WS) potential, the generalised WS potential \cite{yazdankish}, and the Yukawa potential \cite{yukawa}. Yukawa proposed this potential to study the 
interaction between nucleons. The form of Yukawa potential is,
\begin{equation}
    V(r)=-V_{0}\frac{e^{-\alpha r}}{r},
 \end{equation}
where $\alpha$ is the screening parameter. This potential was mainly used to 
get the bound state normalization and energy levels of neutral atoms. Napsuciale and Rodriguez  presented an analytical solution to the 
quantum Yukawa potential \cite{napsuciale}. 

Yukawa potential was combined with linear or other potentials forms and solved 
for the mass spectra of quarkonia (already discussed in section 3.12) \cite{inyang}.

\subsection{The Morse potential}
The Morse oscillator potential has long been used in molecular and nuclear physics to look at the anharmonicities of the vibrational spectra \cite{ahamad}. The Morse barrier potential takes the form,
\begin{equation}
    V(r)=V_{0}\left[2e^{\frac{r}{a}}-e^{\frac{2r}{a}}\right].
\end{equation}
After a finite distance, the potential gives an asymptotically diverging attraction to the outgoing particle while offering a repulsion to an approaching particle at $r<0$.

Jamel studied the heavy quarkonia properties using the trigonometric 
Rosen-Morse potential \cite{jamel}. They have considered heavy quarkonia as a system confined in a hard-wall potential formed by combining a cotangent and 
squared cosecant function. The potential of this combination is trigonometric 
Rosen-Morse potential. The potential was used to examine the state of conformal 
symmetry in the heavy flavor sector. They have obtained the energy eigenvalues and 
eigenfunctions with the help of Nikiforov-Uvarov method. These results have been 
applied to $c \bar c$ and $b \bar b$ quarkonia to obtain the mass spectra and 
root mean square radii. The results showed satisfactory agreement with the available 
experimental and theoretical results.
 
\subsection{The Hulthen potential}
It is one of the short range potentials in Physics \cite{setare}. This 
potential is a modified form of the Eckart potential \cite{eckart}, which has been vastly applied in physics and whose bound-state and scattering 
properties have been studied using different methods. The potential 
has the form,
\begin{equation}
    V(r)=-V_{0}\frac{e^{-\frac{r}{a}}}{1-e^{-\frac{r}{a}}},
 \end{equation}
 
 where $V_{0} = Ze^{2}$ and $a$ is a constant parameter. The Hulthen potential acts like a screened Coulomb potential in short ranges and declines exponentially at large ranges, therefore its 
 bound state capacity is lower than the Coulomb potential. 

Akpan {\it et al} presented the approximate solutions of the 
Schroedinger equation with Hulthen-Hellmann potentials for 
quarkonium systems \cite{akpan}. The equation was solved by Nikiforov-Uvarov method. The wave functions were obtained in the form 
of Laguerre polynomials. The study was able to obtain the mass of 
charmonium and bottomonium states. Quarks were considered to be spinless. The result provides good agreement with the experimental studies and other theoretical studies. A plot of mass spectra with different potential 
parameters was also presented.
 
\subsection{Screened funnel potential}
Screened funnel potential is used for the calculation of $c \bar c$ and 
$b \bar b$ mesons spectra \cite{born, gonzalez}. The potential has the form,

\begin{equation}
    \bar V(r)=\left(\bar \sigma r-\frac{4 \bar \alpha_{s}}{3 r}\right)\left( \frac{1-e^{-\mu r}}{\mu r}\right),
\end{equation}

where $\mu$ is the screening parameter. The $\bar \sigma$ is provided to set it apart from non screening case. The potential will behave like a Coulomb potential at $r\rightarrow0$ and 
at $r\rightarrow\infty$, it will be $\bar\sigma/\mu$. The form of the potential 
is suggested by quenched lattice QCD calculations. The confining part of the potential has the form, 
\begin{equation}
    \bar V(r)_{conf}=\frac{\bar \sigma}{\mu}-\bar \sigma_{r}\frac{e^{-\mu r}}{r}.
\end{equation}
They obtained the masses for low lying hadrons. The model provided a 
rather accurate value for the $b \bar b$ leptonic decay width, masses, and 
radiative decays.

 \subsection{The log potential}
Quigg and Rosner discussed the potential of the 
form \cite{quigg},
\begin{equation}
    V(r)=C \ln \frac{r}{r_{0}},
\end{equation}
with strength $C\sim 3/4$ $GeV$. They showed that quarkonium level spacing becomes independent of quark mass in the 
nonrelativistic limit. They presented features of this potential when it is applied to describe some properties of heavy quarkonium states like energy 
level spacing and leptonic decay width. They have found that the charmonium 
system has a denser spectrum than the modified Coulomb potential. They observed that a $4S$ charmonium level close to $4.25$ $GeV$ was essential for the quark-quark interaction for logarithmic potential.

Machacek and Tomozova discussed the energy spectra and leptonic decay width of the $\Psi$ family with the help of either fractional power or logarithmic 
functions \cite{machacek}. They used the following potentials,
\begin{equation}
 V=Ar^{0.1}+B,
 \end{equation}

 \begin{equation}
      V=A\ln r+B,
\end{equation}

\begin{equation}
     V=A\ln (1+r)+B,
 \end{equation}

 \begin{equation}
     V=A[\ln (1+r)]^{1/2}+B,
  \end{equation}
   \begin{equation}
 V=A[\ln (1+0.2r^{2})]^{1/2}+B,
 \end{equation}
 
where $A$ and $B $ were adjustable parameters to produce a good experimental 
fit. The energy spectrum and leptonic decay widths were obtained. Results 
showed good agreement with the experimental results. However, the third 
potential (eq. 40) exhibited more agreement compared to other potentials. For second 
and third potential, introducing an additional Coulomb term caused the 
effect of reducing effective quark mass for good result.

\subsection{Potential by Bhanot and Rudaz}
Bhanot and Rudaz suggested a new potential for the bound states of heavy quarkonium \cite{bhanot}. The idea of new potential came from the thought that neither a solely Coulombic nor a simply linear potential would be sufficient. The suggested potential takes the form, 
\begin{equation}
    V(r) = \left\{ \begin{array}{lll}
         \frac{-4\alpha_{s}}{3r} & r\leq r_{1} \\
        b \log \frac{r}{r_{0}} & r_{1} \leq r \leq r_{2}.\\
       \frac{r}{a^{2}} & r \geq r_{2}
\end{array} \right.
\end{equation}

This potential has a logarithmic term that interpolates between a linear portion that confines at long distances and a Coulomb term that is asymptotically free at short distances. The study by Quigg and Rosner showed that a logarithmic 
potential for the $Q \bar Q$ could give mass splitting independent of quark mass $m_{Q}$ \cite{quigg}. 
Also, the asymptotic freedom of QCD satisfies the Coulombic and linear terms. 
The independence of $1S-2S$ splitting on $m_{Q}$ for  $Q \bar Q$ suggests a logarithmic part in between. When the three combinations of potentials were selected, the number of model parameters was also increased. They have found 
that the logarithmic part of this potential has a significant role in finding 
the properties of $J/ \Psi$. 
The potential was applied to get the leptonic decay width for the  $J/\Psi$ and $\Upsilon$ family. The results gave excellent agreement on the $\Psi$ spectrum and leptonic decay width. The model could not test for the  $\Upsilon$ leptonic 
decay width due to the lack of experimental data during that time.

\subsection{The Indiana potential} 
 Lichtenberg and Wills studied mass spectra of $\Psi$, 
 $\Upsilon$ and $\zeta$ (bound state of $t \bar t$) family using a 
 quark-antiquark potential named Indiana potential \cite{litchenberg1978}. The potential has the form,
 \begin{equation}
     V= \frac{4\alpha_{0}(1-\frac{r}{r_{0}})^{2}}{3r \ln{\frac{r}{r_{0}}}}+c,
 \end{equation}
 where $\alpha_{0}$ and $r_{0}$ are constants. This is a QCD motivated 
 potential. The need for a logarithmic potential is already discussed by Quigg 
 and Rosner \cite{quigg} and Machacek and Tomozova \cite{machacek}. Indiana 
 potential behaves like a weakened logarithmic potential, 
 $ \frac{1}{r\ln{\frac{r}{r_{0}}}}$ at small distances. At large distances, the 
 potential behaves like $\frac{r}{r_{0}^{2}\ln{r}}$ and the weakening of potential is comparable to 
 linear potential. At $r=r_{0}$ the potential 
 is vanishing. It has one more interesting property, $V({\frac{r_{0}^{2}}{r}})=-V(r)$, implying that the behavior at large and small 
 distances can not be adjusted separately. Spin dependence terms were also 
 added to the potential. The wave function was also obtained to get the 
 leptonic decay widths of vector mesons in this family using Van Royen and 
 Weisskopf formula \cite{van-royen}. The results were compared with the 
 experimental data for $\Psi$ and $\Upsilon$. The leptonic decay width of 
 $\Psi$ showed a decrease with an increase in mass, which did not agree 
 with the experimental results.

\subsection{Potential by Celmaster, Georgi, and Machacek}
 Celmaster, Georgi, and Machacek developed a potential to get $s$- wave meson 
 masses \cite{celmaster},
 \begin{equation}
     V(r)=V_{AF}(r)+V_{INT}(r)+V_{S}(r),
\end{equation} 
 $V_{AF}(r)=\left[\frac{-16\pi}{27}\frac{1}{\ln{\left(\frac{1}{r^{2}\Lambda^{2}e^{2\gamma}}\right)}}+O{\left(\frac{1}{\ln^{3}\left(\frac{1}{r^{2}\Lambda^{2}e^{2\gamma}}\right)}\right)}\right]\frac{1}{r}$, where $\gamma$ is Euler-Mascheroni constant. When $r << \frac{1}{\Lambda}$, they
 expect $V_{AF}(r)$ to dominate. They have also taken into account that energy 
 of the system raises linearly with the increase in separation,  $V_{s}(r)=kr$. $V_{INT}(r)=V_{0}-kre^{-ar}$ was taken as a negligible function 
 compared to the modified Coulombic interaction at short distances and 
 negligible compared to the linear potential at large distances. The hyperfine splitting was taken into account. The parameters in the 
 potential were $a$, $V_{0}$, and quark masses. A fit to the $s$- wave mass 
 spectrum was done by choosing an appropriate quark mass. They demanded 
 that their explanation applies to both light and heavy quark bound states 
 in response to qualitative successes of QCD in describing the light hadron 
 mass spectrum. When they used their model to predict quarks heavier than 
 the charmed quark, the results were qualitatively similar but it was 
 different in detail from those of Eichten and Gottfried's earlier work 
 \cite{eichten2}.
 
 \subsection{Potential by Halzen}
Halzen presented a phenomenological non relativistic quark-antiquark potential 
in the center of mass system. The potential has the form \cite{halzen},
\begin{equation}
    V=V_{c}(r)+V_{d}(r)(\sigma_{i}\cdot \sigma_{j})\\+V_{f}(r)(L \cdot S)+V_{t}(r)\left(\frac{3(\sigma_{i}\cdot r)(\sigma_{j}\cdot r)}{r^{2}}-\sigma_{i}\cdot \sigma_{j}\right),
\end{equation}
where $V_{c}(r)$ is a spherically symmetric infinite potential hole of radius 
$a$, 

\begin{equation}
    \begin{array}{cc}
  V(c)=\Bigg\{ & 
    \begin{array}{cc}
      \infty& r>a \\
       0&     r<a
    \end{array}
\end{array}
,
\end{equation}
here the spin-dependent terms were considered as perturbation. The energy 
eigenvalues were given by the zero point Bessel function. They have obtained the 
energy eigenvalues for $L=0$ and $L=1$ states. The model was also able to 
predict the $D$ wave boson masses. They have also observed that for 
$a=1/1.33m_{\pi}$ and $M_{q}=5GeV/c^{2}$ many other physical properties of 
mesons can be derived. 

\subsection{Potential for $qq \bar q \bar q$ by Weinstein and Isgur}
Weinstein and Isgur considered a nonrelativistic potential model for tetraquarks \cite{weinstein}. This system was already studied with the help of the 
bag model and other relativistic potential models. Those studies concluded that dense discrete spectra exist for these states. Apart from previous potential model studies which were not considering the long range color mixing effects, color confinement forces, and hyperfine interactions were considered here. They have solved the four particle Schroedinger equation. The Hamiltonian was considered as,

\begin{equation}
    H=\sum_{i=1}^{4}\left[m_{i}+\frac{p_{i}^{2}}{2m_{i}}\right]+\sum_{i<j}\left[H_{conf}^{ij}+H_{hyp}^{ij}\right],
\end{equation}

where $H_{conf}$ is harmonic confinement potential and $H_{hyp}$ is color 
hyperfine interaction. The Hamiltonian did not consider the anharmonicities, 
the effect of possible $q \bar q$ annihilations via gluons and some relatively 
small spin-orbit and tensor effects. The model did not give evidence for any denser discrete spectrum for 
the states. The model confirmed that light $qq \bar q \bar q$ states can 
exist with $0^{++}$. Also, it allowed the existence of meson-meson bound states like the 
nucleon-nucleon interaction of deuteron.

\subsection{Potential by Gupta, Repko, and Suchyta}
 Gupta, Repko, and Suchyta developed a nonsingular potential model to 
 investigate the quarkonium spectra \cite{gupta1}. They have used the 
 semirelativistic Hamiltonian of the form,

 \begin{equation}
     H=2(m^{2}+p^{2})^{1/2}+V_{p}(r)+V_{c}(r),
 \end{equation}
 
 where $V_{p}$ and $V_{c}$ are the perturbative and confining potentials. For 
 confining potential they made use of a mixed scalar and vector exchange 
 potential. The form of $V_{c}$ is given by, 
 
  \begin{equation}
 V_{c}=Ar+\frac{C_{1}}{m^{2}r}(1-e^{-2mr}){S_{1}\cdot S_{2}}+\frac{C_{2}}{m^{2}r}(1-\frac{1}{2}f_{1}){L\cdot S}+\frac{C_{3}}{m^{2}r}(1-\frac{3}{2}f_{2})S_{12},
  \end{equation}
  
here $C_{1}$, $C_{2}$ and $C_{3}$ are arbitrary constants. They have obtained the energy levels, leptonic decay widths, and $E_{1}$ transition width. The values agreed well with experimental data of $b \bar b $ and reasonably good for $c \bar c$.

\subsection{The Richardson potential}
Richardson proposed a potential incorporating the asymptotic freedom and linear quark confinement of QCD \cite{richardson}. The potential 
generates the spectrum of the triplet $ c \bar c$ and the triplet $ b \bar b$. 
The form of the potential is,
\begin{equation}
    \tilde{V}(q^{2})=-\frac{4}{3}\frac{12\pi}{33-2n_{f}}\frac{1}{q^{2}}\frac{1}{\ln\left({1+\frac{q^{2}}{\Lambda^{2}}}\right)},
\end{equation}

where the only parameters in the model are scale size $\Lambda$ and the quark 
masses. The value of $n_{f}$ was chosen as three, assuming the effect of heavy quark will be 
negligible at a distance they were studying. Here the Fourier transform for the coordinate space potential $V(r)$ was taken by considering the one gluon 
exchange amplitude, which is proportional to $\tilde{V}(q^{2})$. They have not 
considered spin-dependent effects to the potential. For $\Upsilon$ and $\Psi$ 
system, experimental results showed a reasonably good agreement with the model.

Bagchi {\it et al} used Richardson potential to study energies and 
magnetic moments of $\Omega^{-}$ and $\Delta^{++}$ \cite{bagchi}. They have 
modified the potential with a new set of scale parameter values 
for asymptotic freedom and confinement. Moreover, they expect this potential 
can be a good base for studying the baryon properties.

\subsection{Klein-Gordon (KG) oscillator potential}
Grunfeld and Rocca presented a relativistic confining potential using the Klein-Gordon oscillator to get the mass spectra of $c \bar c$ and 
$b \bar b$ \cite{grunfeld}. The KG oscillator was introduced by Bruce and 
Minning \cite{bruce}. This oscillator will behave like a harmonic oscillator 
(HO) in the non relativistic limit. The two body problem was solved to obtain 
the mass spectra. A KG equation takes the form \cite{bruce},
\begin{equation}
    -\frac{\partial^{2}}{\partial t^{2}}\Psi(\mathbf{q},t) = 
\left (\mathbf{p^{2}}+m^{2}\mathbf{q} \cdot \hat{\Omega}^{2} \cdot \mathbf{q}+m \hat{\gamma} (tr \Omega)+m^{2} \right)\Psi(\mathbf{q},t),
\end{equation}
where $\hat{\Omega}$ is a $3 \times 3$ matrix and $\omega_{i}$ is the 
oscillator frequency.

\begin{equation}
    \hat{\Omega}_{ij}=\omega_{i}\delta_{ij}.
\end{equation}

The quark mass and $\omega$ are the two free parameters. The quarks were 
considered as spinless. The results were compared with the Klein-Gordon equation 
with linear and quadratic potentials \cite{kang, ram}. They have also compared their values with 
a four dimensional harmonic oscillator model in a quantum relativistic frame \cite{arshansky}.
The results have shown a good agreement with the theoretical and experimental data.

\section{Lattice QCD (LQCD)}
At high energies, perturbation theory can be used to get the 
analytical solutions of QCD. However, the perturbation method fails at lower 
energies. Therefore, an alternative approach, the Lattice Quantum 
Chromodynamics (LQCD) is used to calculate the QCD predictions numerically. 
The domain in which the perturbation method fails LQCD provides a 
nonperturbative tool for finding the hadron spectrum and the matrix elements. 
LQCD is developed on a discrete Euclidean space-time grid and retains the 
fundamental characters of QCD. Field theory is applied to LQCD via the Feynman path integral method. Numerical simulations of LQCD use Monte-Carlo integration 
of the Euclidean path integral \cite{gupta}. 

LQCD has two applications. Lattice regularisation acts as a nonperturbative regularisation scheme and can be used to perform any typical perturbative calculations. Second, using methods similar to those utilized in statistical mechanics systems, it is possible to simulate LQCD on a computer by converting QCD into a space-time lattice. The correlation functions of hadronic operators and matrix elements of any operator between hadronic states can be calculated using these simulations in terms of the fundamental quark and gluon degrees of freedom. The chiral symmetry breaking, equilibrium properties, and confinement mechanisms of QCD at finite temperatures can also be addressed by LQCD. It offers a useful function where the input settings can be fixed. Therefore, it is possible to estimate quark masses and the strong coupling constant $\alpha_{s}$. These facts can be utilised to constrain theories like phenomenological models, heavy quark effective theory and chiral perturbation theory. Testing QCD theories and processes with significant momentum transfers is the primary aim of LQCD.

LQCD studies are helpful in getting new hadron states. Tetraquark states $X(3872)$ \cite{chiu2}, $Y(4260)$ \cite{lacock}, the charged $Z_{c}$ states \cite{leskovec}, the doubly heavy tetraquark states \cite{padmanath}, and the hidden-charm pentaquark states \cite{skerbis} were studied with the help of LQCD. The mass spectra 
of the tetraquark state $Y(4260)$ were studied in quenched lattice QCD with 
exact chiral symmetry \cite{chiu}. The mass spectra of hybrid charmonium ($c \bar c g$), molecular operator, and diquark-antidiquark operators were 
computed. It has also suggested a possibility that $Y(4260)$ can be an excited state of $c \bar c$. 

Brambilla, Consoli and  Prosperi gave a derivation for the quark-antiquark 
potential in a Wilson loop context \cite{brambilla}. The basic assumptions, 
the condition used for the validity of potential and the relation with the 
flux tube model were considered in the Wilson-loop approach. The potential 
contains three terms. One is a static term (stat), the spin-dependent term (SD) and the velocity-dependent term (VD),

\begin{equation}
    V^{q \bar q}=V^{q \bar q}_{stat}+V^{q \bar q}_{VD}+V^{q \bar q}_{SD}.
\end{equation}

The same approach was extended to get the three quark potential also,

\begin{equation}
V^{3q}=V^{ 3q}_{stat}+V^{3 q}_{VD}+V^{3q}_{SD}.
    \end{equation}
    
They have presented the form of each term. In the case of three quark 
potential, they have observed that the short-range part of the equation 
for the three quark potential was a pure two-body type potential. This can 
be compared with the electromagnetic potential for three charges. The spin-dependent term contains a 
long-range part which was coinciding with the expression given by Ford as \cite{ford},

\begin{equation}
\sigma(r_{1}+r_{2}+r_{3})\beta_{1}\beta_{2}\beta_{3}.
\end{equation}

The spin-dependent potential for three quark has been consistent with the 
Wilson loop context. The order $1/m^{2}$ for $qqq$ potential was also new to 
their study. They have done spin-independent relativistic correction on 
$q \bar q $ and $qqq$. This way, better results were obtained by assuming scalar 
confinement.

Bicudo {\it et al} gave a theoretical method to get the mass and decay width of doubly heavy tetraquark $u d \bar b \bar b$ \cite{bicudo}. The potential between two heavy 
antiquarks [$ \bar Q \bar Q$] and two light quarks [$qq$] was parametrized by a 
screened Coulomb potential using lattice QCD,

\begin{equation}
V(r)=\frac{-\alpha}{r}e^{\frac{-r^{2}}{d^{2}}},
\end{equation}

where $\alpha$ and $d$ were parameters dependent on isospin and angular 
momentum of $qq$ pair. The Schroedinger equation was solved for the potential 
to obtain mass and decay width. Mass for the state was obtained as 
$m=10576\pm4$ $MeV$ and $\Gamma=112^{+10}_{-103}$ $MeV$.

\section{Summary and outlook}
From the above review, we have seen that hadron spectroscopy can be studied by 
many methods. In this article, we have classified them on models and potentials. 

Table \ref{tab:Table 1} shows the list of different models we have reviewed and particles studied in these models. BSE formalism, CMI method, and QCD sum 
rules have studied fully heavy pentaquarks. However, such type of system is yet
to be detected experimentally. In BSE and CMI models fully heavy tetraquarks 
are studied. Doubly heavy and triply heavy tetraquarks are also discussed in 
different models successfully. Such tetraquarks have been observed recently 
also. We expect that this work will be helpful in explaining these data. QCD 
sum rule is also found to be helpful in the study of hexaquarks. Unfortunately, 
most of the studies are devoted to mass calculations, only a few studies have 
investigated the decay properties.% compared to mass spectra. 

The CMI Hamiltonian is useful in determining the mass and decay properties of 
fully heavy pentaquarks. However, the determination of parameters, $m_{i}$ and 
$v_{ij}$ in the CMI Hamiltonian is difficult due to the lack of sufficient 
experimental data. As we have observed, some models consider exotics as 
diquarks. It is expected that the diquark correlation can answer some 
puzzles in exotic hadron spectroscopy, including the rarity of exotics in QCD. 

\begin{table}
\FloatBarrier
\caption{\label{arttype}Different models and the system studied in that model.}
\footnotesize\rm
\begin{tabular*}{\textwidth}{@{}l*{15}{@{\extracolsep{0pt plus10pt}}l}}
\br
Model&System studied\\
\hline
Bag model&$q \bar q$, $qqq$, $qq \bar q \bar q$\\
\hline
BSE&$q \bar q$, $q\bar Q $, $Q \bar Q$, $bcq$, $Q \bar QQ\bar Q$, $QQQQ \bar Q$\\
\hline
CMI&$QQq$, $QQQ$, $Q  Q \bar Q\bar q$, $Q  Q \bar Q\bar Q$, $QQQQ \bar Q$\\
\hline
QCD sum rule&$q \bar q$, $QQq$, $QQQQ \bar Q$, $QQQQQQ$\\
\hline
Diquarks&$qc \bar q \bar c$, $sc \bar s \bar c$\\
\hline
Skyrme model&$qqq$&\\
\hline
Hadroquarkoium &$qq \bar Q \bar Q$, $qqqQ \bar Q$&\\
\hline
OPEP &$P_{c}$ pentaquark\\
\br
\label{tab:Table 1}
\end{tabular*}
\FloatBarrier
\end{table}

\begin{figure}[htbp]
\caption{Plot of different potentials.}
\centerline{\includegraphics[width=10cm]{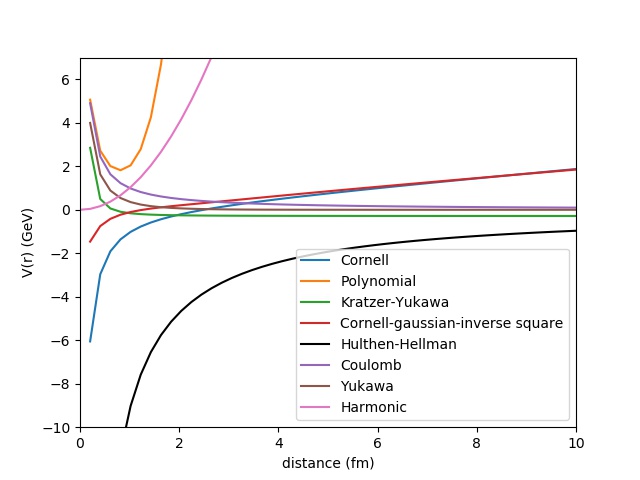}}
\label{fig: Figure 1}
\end{figure}

We have plotted the form of five phenomenological potentials (Cornell, 
polynomial, Kratzer-Yukawa, Cornell-Gaussian-inverse square, and 
Hulthen-Hellman) given in Figure \ref{fig: Figure 1}. The mathematical form 
of these potentials is given in Table \ref{tab:Table 2}. Apart from these, 
Coulomb potential, Harmonic potential, and Yukawa potential are also included 
in the plot. All the potentials are used to get the mass spectra 
of heavy quarkonia ($c \bar c$ and $b \bar b$). Cornell potential and 
combination of Cornell, Gaussian, and inverse square potentials show almost 
same behavior after $2fm$ separation. Hulthen-Hellman potential shows a large 
deviation from other potentials. However, potential approaches to other 
potential after $8fm$. The shift in the potential is evident in Kratzer plus 
Yukawa potential compared to Yukawa potential. Similarly, we can identify that 
the potential is modified when a linear term is added to Coulomb potential, 
which is Cornell potential. 

                                                                                From Table \ref{tab:Table 2} we can see that phenomenological potential
study is quite successful in heavy quarkonium systems. For small quark 
separations ($\sim 4fm$) the nature of most of the potentials are similar 
(Figure \ref{fig: Figure 1}) therefore, the mass spectra estimated are also 
showing similar pattern. Therefore, we expect that the decay properties 
estimations should also be nearly the same. However, very limited work has 
been done so far for heavy tetra and pentaquarks.

We have compared the mass spectra of a few states of charmonium in Table 
\ref{tab:Table 3} and bottomonium in Table \ref{tab:Table 4} (without spin 
correction) in some potentials. For $1S$, all potentials approach gave the same 
result ($9.460$ $GeV$ for bottomonium and $3.096$ $GeV$ for charmonium, which 
is not shown in the table). Mass spectra do not show much deviation for $2S$ 
and $1P$ states. 

\begin{table}
\caption{\label{arttype}Different potentials, it's mathematical form and the system studied.}
\footnotesize\rm
\begin{tabular*}{\textwidth}{@{}l*{15}{@{\extracolsep{0pt plus12pt}}l}}
\br
Potential& $V(r)$&System studied\\
\hline
Cornell&$-\frac{a}{r}+br$&$c \bar c, b \bar b$, $c \bar c q \bar q$, $b b \bar b \bar b$,$ b \bar q b \bar q$\\
\hline
Polynomial& $\frac{b}{r}+ar+dr^{2}+pr^{4}$ &\vspace{0.05cm} $c \bar c, b \bar b$
\\
\hline
Cornell, Gaussian and inverse square &$\frac{a}{r}+dr+\frac{b}{r^{2}}+c_{0}e^{\frac{-\alpha^{2}r^{2}}{2}}$  & $B$ and $D$ mesons\\
\hline
Kratzer and screened Coulomb&$-\frac{b}{r}+\frac{c}{r^{2}}+\frac{pe^{-ar}}{r}+a$ &$c \bar c, b \bar b  $\\
\hline
Yukawa, linear and harmonic &  $ar+dr^{2}-\frac{b e^{-cr}}{r}$&$c \bar c, b \bar b, \bar b c$\\
\br
\label{tab:Table 2}
\end{tabular*}
\end{table}

\begin{table}
\caption{\label{arttype} Mass spectra for charmonium states in different potentials (in $GeV$).}
\footnotesize\rm
\begin{tabular*}{\textwidth}{@{}l*{15}{@{\extracolsep{0pt plus12pt}}l}}
\br
$V(r)$&Constants & $2S$ &$1P$&$1D$\\
\hline
$ar-\frac{b}{r}$ \cite{kuchin}& \tiny{{$a=0.2$$GeV^{2}$, $b=1.244$} } &  3.686 & 3.225&3.504\\
 \hline
$\small{{ar^{2}+br-\frac{c}{r}+\frac{d}{r^{2}}}}$ \cite{omugbe}&  \tiny{{$a=0$, $d=0,$ $b=0.202GeV^{2}$, $c=1.664$}} &3.689   & 3.262&\ 3.515 \\
\hline
$a-\frac{b}{r}+\frac{c}{r^{2}}+\frac{pe^{-ar}}{r} \cite{inyang}$ & \tiny{{$a=-0.2860$$GeV$, $b=0.001GeV$,  $c=0.1306GeV$, $p=0.0022GeV$}} &3.686   & 3.295  & 3.583 \\
\hline
$\frac{b}{r}+ar+dr^{2}+pr^{4}$ \cite{mansour2}&  \tiny{{$a=10.7$$GeV$,\newline $b=6.39286GeV$, \newline $d=-0.495eV$,\newline$p=7.1GeV$}}  &-  & - &3.6861 \\  
\hline
$\frac{-a_{0}e^{-ar}}{1-e^{-ar}}-\frac{b}{r}+\frac{ce^{-ar}}{r}$ \cite{akpan}& \tiny{{$a_{0}=-1.591$$GeV$, $b=9.649GeV$, $c=0.028$}}  &3.686   &3.521& 3.768  \\
\br
\label{tab:Table 3}
\end{tabular*}
\end{table}

\begin{table}
\caption{\label{arttype}Mass spectra for bottomonium states in different potentials (in $GeV$).}
\footnotesize\rm
\begin{tabular*}{\textwidth}{@{}l*{15}{@{\extracolsep{0pt plus12pt}}l}}
\br
{$V(r)$ }&Constants &$2S$& $1P$ &$1D$  \\
\hline
$ar-\frac{b}{r}$ \cite{kuchin}&  \tiny{{$a=0.2$$GeV^{2}$, $b=1.569$} }  & 10.023 &9.691& 9.864\\
 \hline
$\small{{ar^{2}+br-\frac{c}{r}+\frac{d}{r^{2}}}}$\cite{omugbe}&  \tiny{{$a=0$,$d=0$ $b=0.202GeV^{2}$, $c=1.664$}}  &10.023    & 9.608&9.814 \\
\hline
$a-\frac{b}{r}+\frac{c}{r^{2}}+\frac{pe^{-ar}}{r} \cite{inyang}$ &\tiny{{$a=-0.0723$$GeV$, $b=0.001GeV$, $c=0.050GeV$, $p=0.0022GeV$}} &10.569    & 9.661   &9.943  \\
\hline
$\frac{-a_{0}e^{-ar}}{1-e^{-ar}}-\frac{b}{r}+\frac{ce^{-ar}}{r}$ \cite{akpan} &  \tiny{{$a_{0}=-1.591$$GeV$, $b=9.649GeV$, $c=0.028$}}  &10.023   &9.861&    10.143       \\
\br
\label{tab:Table 4}
\end{tabular*}
\end{table}

Lattice QCD has explained some of the $XYZ$ tetraquark states and large momentum
transfers. However, quantitative confirmation is still needed. The predictions 
provided by lattice QCD are reliable only for hadrons with heavy quarks. 

Therefore, we see that significant work is done on hadron spectroscopy of 
tetra and pentaquark systems in the framework of different potentials and 
models. But most of the works are concentrated for mass calculations. Decay 
study of such systems are also very important and required. Still a lot 
of work is expected in this area.
\newpage
\section*{Acknowledgements}
We are grateful to Manipal Academy of Higher Education for the financial 
support under Dr. T. M. A. Pai scholarship program.

\end{document}